# Influence of High-Speed Railway System on Inter-city Travel Behavior in Vietnam

Tho V. LE [a], Junyi ZHANG [b], Makoto CHIKARAISHI [c], Akimasa FUJIWARA [d]

[a] *Department of Transportation and Economics, University of Transport and Communications, Vietnam; E-mail: letho.utc@gmail.com*
[b,c,d] *Graduate School for International Development and Cooperation, Hiroshima University, Japan*
[b] *E-mail: zjy@hiroshima-u.ac.jp*
[c] *E-mail: chikaraishim@hiroshima-u.ac.jp*
[d] *E-mail: afujiw@hiroshima-u.ac.jp*

**Abstract**: To analyze the influence of introducing the High Speed Railway (HSR) system on business and non-business travel behavior, this study develops an integrated inter-city travel demand model to represent trip generation, destination choice, and travel mode choice behavior. The accessibility calculated from the RP/SP (Revealed Preference/Stated Preference) combined nested logit model of destination and mode choices is used as an explanatory variable in the trip frequency models. One of the important findings is that additional travel would be induced by introducing HSR. Our simulation analyses also reveal that HSR and conventional airlines will be the main modes for middle distances and long distances, respectively. The development of zones may highly influence the destination choices for business purposes, while prices of HSR and Low-Cost Carriers affect choices for non-business purposes. Finally, the research reveals that people on non-business trips are more sensitive to changes in travel time, travel cost and regional attributes than people on business trips.

## 1. INTRODUCTION

The innovation of a high-speed railway (HSR) system to connect cities was first realized in Japan, known as the Shinkansen (or bullet train), in 1964. Since then, the HSR system has been widely introduced in many countries such as France, Germany, Spain, Italy, Taiwan, and China. Recently, some other countries are also showing interest in the HSR, including the USA and Vietnam. It is believed that HSR will induce travel demand, both in short-run effects (e.g., route switches, mode switches, changes of destination, and new trip generation) and long-term effects (e.g., change in household auto ownership and spatial re-allocation of business facilities). Drawing on the insights of existing travel behavior studies, it is expected that the demand for HSR will be highly affected by the quality of service of other travel modes, such as cars, buses, conventional trains, and airlines (Yao and Morikawa 2005).

   Some researchers have evaluated the potential impacts of HSR on economics, regions, environment, and travel behavior (Givoni and Dobruszkes 2013, Albalate and Bel 2012; USCM 2010; Givoni 2006). Another hot topic in the context of inter-city transportation systems is the emergence of Low-Cost Carrier (LCC) airlines (Noran et al. 2001; Burghouwt et al. 2003; Dennis 2004; Fan 2006).

   The demand for inter-city transportation in Vietnam has been increasing year by year because of economic development and population growth. However, the more and more serious traffic congestion and accidents have negatively influenced regional economic development, national productivity and competitiveness, and environmental quality.

To resolve the prevalent problems and prepare necessary infrastructure for meeting the future travel demand, recently the Vietnamese government has been considering the construction of a HSR line in the near future to connect the two biggest cities in Vietnam, Hanoi in the north and Ho Chi Minh City in the south. The construction is expected to have strong potential influences on economics, land use, and travel behavior in the North-South corridor of Vietnam.

Furthermore, some companies have been operating LCCs serving some main destinations in Vietnam in a recent decade. Therefore, LCCs and HSR will be competing with each other, suggesting that HSR services, such as price, frequency, travel time, and punctuality, should be carefully designed. However, to the authors' knowledge, there is little quantitative analysis on the competition between LCCs and HSR and factors affecting customers' decisions, though there are a number of studies focusing on the competition between LCCs and conventional carriers.

The differences between HSR and LCCs in terms of the level of services can be summarized as follows. First, HSR stations can easily be located near city centers, cancellations and delays are rare, and the service is punctual, compared with LCCs. On the other hand, even though the access time to airport is usually longer than that to HSR station, in-vehicle travel time of LCCs is considerably shorter than that of a HSR since airport is typically located in suburban area. In Vietnam, the HSR system is planned to be constructed with double tracks for the entire distance of 1,570 km from Hanoi to Ho Chi Minh City. In general, air travel becomes dominant when the distance is more than 1,000 km as we will discuss later in details. Therefore, the demand for the HSR system should be carefully estimated to evaluate the feasibility of the HSR project.

The lessons from Taiwan's experience in planning and operating of HSR system may be vital for implementing the HSR project in Vietnam. There were several issues, which led to the actual ridership of their HSR system being less than expected. These issues include accessibility to HSR stations and integration between modes of transport (Yung-Hsiang Cheng 2009). The current solution is to provide free shuttle buses to carry passengers to and from city centers to HSR stations due to their location in peripheral areas. In line with that, the Taiwanese government has proposed a system to integrate HSR with bus, airline, and conventional rail (Taiwan High Speed Rail 2016).

Albalate and Bel (2012) studied HSR in five countries, namely Japan, France, Germany, Spain, and Italy, and found that HSR has positive impacts on factors such as mobility, environment, economic, and regional development. Travel demand was found as the core factor in the decision to construct the HSR lines and the decision to determine the opening or commercial operation date. According to the authors' best knowledge, there is no study at the micro-level about the effect of introducing HSR on inter-city travel demand in the context of a developing country such as Vietnam. This study attempts to fill in this gap by developing an integrated inter-city travel demand model representing trip generation, destination choice, and travel model choice behavior. A comprehensive questionnaire survey was designed conducted. We also determine the influencing factors such as level of service of travel modes, regional characteristics, and social demographic characteristics on the decision of mode choice, destination choice, and trip generation.

The rest of the paper is organized as follows. Section 2 further reviews existing studies on the impacts of HSR on inter-city travel demand. Section 3 introduces the integrated inter-city travel demand model developed in this study, followed by data in Section 4. Section 5 discusses the influential factors on inter-city travel demand based on the model estimation results. We also conduct simulation analysis to identity the relevant policy impacts. Major findings and some future tasks are summarized in Section 6.

## 2. LITERATURE REVIEW

Since the first introduction of the Shinkansen in Japan in 1964, there are numerous countries all over the world, which have constructed and proposed HSR systems. With its advantages, HSR can influence mobility, environment, economic and regional development along the HSR routes.

Albalate and Bel (2012) studied data provided by the European Commission (1996) and showed the changes in modal shares following the introduction of HSR in several routes in Europe. In the Paris-Lyon (France) line, between 1981 and 1984, the modal share of air traffic fell from 31% to 7%, and that of car and bus traffic slightly decreased by 8% from 29%, while rail traffic rose from 40% to 72%. The similar modal split trend can be observed between 1991 and 1994 in the Madrid-Seville line in Spain. At that time frame, the modal share of air traffic fell from 40% to 13%, and that of car and bus from 44% to 36%, whereas train travel increased by 35% to 51%. The data proves that the introduction of an HSR line can alter the modal split between two cities, and modal shares are subject to dramatic changes.

It is reported by King (1996) that the HSR systems in Japan and France have produced as high as 35% of additional induced travel and about 30% for modal shifts (as mentioned in Yao and Morikawa 2005). It is believed that the opening of HSR will induce travel demand, both in short run effects (e.g., route switches, mode switches, changes of destination, and new trip generation) and long term effects (e.g., change in household auto ownership and spatial reallocation of activities). For the Tokyo-Nagoya-Osaka corridor (about 500km), the induced travel accounts for 16.5% of travel demand measured in vehicle miles traveled and 14.5% of travel demand measured in number of trips. In addition, the demand for HSR may be affected by the quality of service of other modes, especially, if LCC is available. Cascetta and Coppola (2014) show that the levels of services (travel time, travel cost, and service frequency) positively influence trip frequency. In addition, the paper confirmed that the elasticity of non-business trips is higher than that of business trips, especially for shorter distances.

The study of Cascetta and Coppola (2012) employed the both RP and SP data for the estimation of elastic demand by schedule-based multimodal assignment model. The research introduced a model system to forecast the national passenger demand for multiple scenarios of macroeconomic, transport supply, and HSR market. The higher degrees of substitutions among specific subsets of mode/service/run alternatives have been captured from the developed model.

By investigating the HSR systems in Japan, France, Germany, Spain, and Italy, Albalate and Bel (2012) confirmed that besides business trips, tourism is the first sector to show an immediate effect following the inauguration of an HSR line. The changing of destination choice in non-business trips is mainly based on tourism activity due to the location of visiting friends or relatives being fixed, while there are a variety of places for travelling purposes. Tourism activity is affected by some factors, such as tourism resources and facilities, motivation and characteristics of travelers, and accessibility to the destination. The first two factors are difficult to manage, but the later can be controlled through transportation policies as they can set up or modify the level of services of transportation modes.

The evidence from Levinson's (2012) research, on the other hand, confirmed the two opposite effects of HSR: positive accessibility benefits in metro areas served by stations, and negative nuisance effects along the lines themselves. People and business who are located near the lines but use HSR less frequently may be little benefit, while HSR system still brings widespread advantages for metropolitan areas. High-speed lines, however, have much higher nuisance effects than local transit.

In the context of the transportation market including HSR and LCC, there is no research on integrated model for tourism travel demand to the best knowledge of the authors. There are some studies dealing with tourist behavior, which directly or indirectly refer to the availability of HSR. The research of Masson and Petiot (2009) stated the possible effect of HSR between Perpignana (France) and Barcelona (Spain) on the reinforcement of tourism attractiveness by using a core-periphery model. The HSR could reinforce the agglomeration of the tourism industry on Barcelona, which is a more developed area than Perpignana. This study is concerned about tourism destination development, but not about modeling for tourism travel demand. Moreover, the study of Albalate and Bel (2012) reviewed HSR in Japan and four European countries, and found that tourism and the service sector benefit the most from the construction of HSR.

Some researchers prove that HSR is the dominant mode in distances less than 800 km, while airlines are the dominant mode in distances of more than 1,000 km (MLIT, Japan. 2010, Albalate and Bel 2012, Shiomi, 1999). However, in those studies, LCC was absent in the transportation market. In addition, HSR has several advantages, such as easy to access and egress to stations that are usually located in the urban center, very rare cancellations and delays, and high punctuality. Even though the LCC fare is cheaper than that of HSR, the tickets need to be booked in advance, and there are also higher rates for cancellations and delays. When the difference in travel time between HSR and airlines is not significant, it can be supposed that the mode share will highly depend on government strategies, such as policies for setting the level of service and the development priority for which kind of mode.

In addition, the extensive study on HSR of Albalate and Bel (2012) found that modal distribution of traffic is affected when HSR start operation with greatest impact on the airline industry. The study of Park and Ha (2006) suggests that for the Paris-Lyon route (450km), the share of air transport decreased by half, from 30% to 15%. However, with the longer distance like the Paris-Marseilles route (700 km), the share of airline dropped from 45–55% to 35–45% and the Paris-Nice route (900 km) dropped from 55–65% to 50–60%. This implies that the competition for air transport and HSR is highly influenced by travel distances. Decrease of airline modal share is more significant for short and middle trip distances than that of long trip distances. Delaplace and Dobruszket (2015) studied the competition of low cost HSR with traditional HSR and LCC. They found fare is the main influence on mode chooses. In addition, Chantruthai, et al. (2014) found travel time, fare, user's occupation, household income, education level, and trip purposes significantly influence on HSR and LCC mode choice behavior.

It is common in literature to combine RP/SP data in order to exploit the advantages of each data source. The advantages of RP/SP combined approach are (1) the efficient use of data (multiple data sources are used), (2) correcting biases, which potentially exist in SP data, and (3) identifying the effects of new services (Morikawa, 1994). In this study, following Morikawa (1994), a RP/SP combined travel demand model is developed to explore the impacts of trip characteristics and the level of services on mode choice, destination choice behavior, and trip generation in Vietnam, where LCC plays as an important inter-city travel mode.

## 3. INTEGRATED INTER-CITY TRAVEL DEMAND MODEL

In the context of modeling inter-city travel, Koppelman (1989), Zhuang et al. (2007) and Yao and Morikawa (2005) revealed that inter-city travel behavior components, such as trip generation, destination choice, travel mode choice, and route choice are interrelated and should be modelled in an integrated framework. In line with this, this study develops a nested inter-city travel demand model as shown in Figure 1, where trip generation and frequency, destination choice, and the mode choice are modeled in a hierarchical manner. The feedback mechanism represents the interrelationships of the travel choices.

Our modeling system consists of two parts: The first part is the joint model of travel mode and destination choices based on the nested RP/SP combined model. Note that business trips and non-business trips are modeled separately, since these two decisions may be substantially different for example due to differences in schedule constraints. The second part is the trip generation model, where the accessibility index, which is calculated based on the estimation results of mode and destination choice models, is added as an explanatory variable. The detailed model specification is given below.

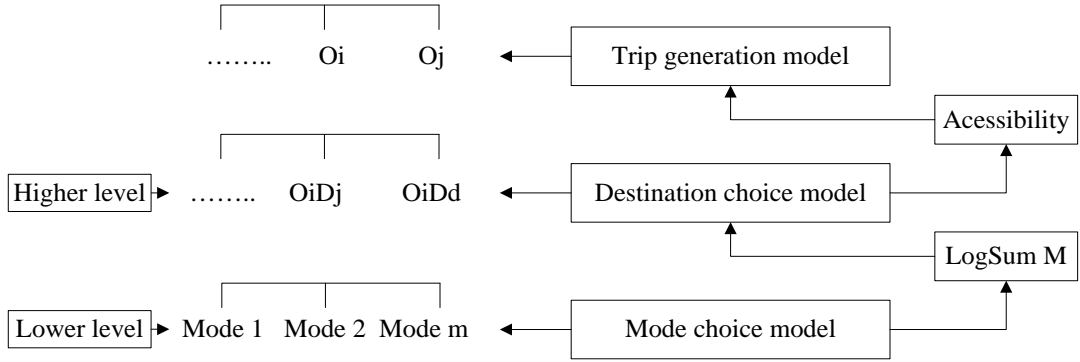

Figure 1 The nested structure of inter-city travel demand model
(revised from Koppelman (1989) and Yao and Morikawa (2005))

*3.1 The RP/SP Combined Destination and Mode Choice Model*

In the RP/SP combined destination and mode choice model, we assume that both revealed and stated preferences can be modeled by random utility models. Concretely, the utility functions for RP and SP situations are assumed as follows:

$$U_d^{RP} = V_d^{RP} + \varepsilon_d^{RP} \quad (1)$$
$$U_d^{SP} = V_d^{SP} + \varepsilon_d^{SP} \quad (2)$$

where $V_d^{RP}$ and $V_d^{SP}$ are the observable components of the utilities for choosing destination $d$ in RP and SP situations respectively, and $\varepsilon_d^{RP}$ and $\varepsilon_d^{SP}$ are the error components of the utilities for choosing destination $d$ in RP and SP situations respectively.

Similarly, the utility functions of mode choice models are defined as follows:

$$U_m^{RP} = V_m^{RP} + \varepsilon_m^{RP} \quad (\text{mode } m \in M^{RP}) \quad (3)$$
$$U_m^{SP} = V_m^{SP} + \varepsilon_m^{SP} \quad (\text{mode } m \in M^{SP}) \quad (4)$$

where $V_m^{RP}$ and $V_m^{SP}$ are the observable components, $\varepsilon_m^{RP}$ and $\varepsilon_m^{SP}$ are the error components in RP and SP situations respectively, and $M^{RP}$ and $M^{SP}$ are the available alternatives in RP and SP choice contexts respectively.

We further specify the observable components of the utilities $V_d^{RP}$ and $V_d^{SP}$ as follows:

$$V_d^{RP} = \lambda C_d^{RP} + \theta_d^{RP} \Gamma_d^{RP} \quad (5)$$

$$V_d^{SP} = \varphi Q_d^{SP} + \theta_d^{SP} \Gamma_d^{SP} \tag{6}$$

where $C_d^{RP}$ and $Q_d^{SP}$ is vectors of explanatory variables, $\lambda$ and $\varphi$ are vectors of unknown parameters to be estimated, $\theta_d^{RP}$ and $\theta_d^{SP}$ are the logsum parameters associated with destination $d$, and $\Gamma_d^{RP}$ and $\Gamma_d^{SP}$ are the logsum variables (or inclusive values) which are defined based on a set of available travel modes' levels of services from an origin to destination $d$. We further specify $\theta_d^{RP}$, $\theta_d^{SP}$, $\Gamma_d^{RP}$ and $\Gamma_d^{SP}$ as follows:

$$\theta_d^{RP} = \frac{\exp(\omega k^{RP})}{1 + \exp(\omega k^{RP})} \tag{7}$$

$$\Gamma_d^{RP} = \ln \sum_{m \in M_d^{RP}} \exp \frac{V_m^{RP}}{\theta_d^{RP}} \tag{8}$$

$$\theta_d^{SP} = \frac{\exp(\omega k^{SP})}{1 + \exp(\omega k^{SP})} \tag{9}$$

$$\Gamma_d^{SP} = \ln \sum_{m \in M_d^{SP}} \exp \frac{V_m^{SP}}{\theta_d^{SP}} \tag{10}$$

where $k^{RP}$ and $k^{SP}$ are vectors of explanatory variables, and $\omega$ is a vector of unknown parameters. We also assume $V_m^{RP}$ and $V_m^{SP}$ are defined as:

$$V_m^{RP} = \beta' X_m^{RP} + \alpha' W_m^{RP} \tag{11}$$
$$V_m^{SP} = \mu(\beta' X_m^{SP} + \gamma' Z_m^{SP}) \tag{12}$$

where $X_m^{RP}$ and $X_m^{SP}$ are the common attribute vectors; and $W_m^{RP}$ and $Z_m^{SP}$ are the specific attribute vectors for the RP and SP data, respectively. $\mu$ is the scale parameter which adjusts the differences in standard deviations of error components between RP and SP models.

The joint probabilities of individual's RP and SP choices can be described as:

$$P_{nm}^{RP}(d) = P_{nm|d}^{RP} P_{nd}^{RP} \tag{13}$$
$$P_{nm}^{SP}(d) = P_{nm|d}^{SP} P_{nd}^{SP} \tag{14}$$

where $P_{nd}^{RP}$ and $P_{nd}^{SP}$ are the probabilities of choosing destination $d$ in RP and SP choice contexts respectively, and $P_{nm|d}^{RP}$ and $P_{nm|d}^{SP}$ is the conditional probabilities of choosing travel mode $m$ given destination $d$ in RP and SP choice contexts respectively. Concretely, $P_{nd}^{RP}$, $P_{nd}^{SP}$, $P_{nm|d}^{RP}$ and $P_{nm|d}^{SP}$ are defined as follows:

$$P_{nd}^{RP} = \frac{\exp(V_d^{RP})}{\sum_{d' \in D^{RP}} \exp(V_{d'}^{RP})} \tag{15}$$

$$P_{nm|d}^{RP} = \frac{\exp\left(\frac{V_m^{RP}}{\theta_d^{RP}}\right)}{\sum_{m' \in M^{RP}} \exp\left(\frac{V_m^{RP}}{\theta_d^{RP}}\right)} \tag{16}$$

$$P_{nd}^{SP} = \frac{\exp(V_d^{SP})}{\sum_{d' \in D^{SP}} \exp(V_{d'}^{SP})} \tag{17}$$

$$P_{nm|d}^{SP} = \frac{\exp(\frac{V_m^{SP}}{\theta_d^{SP}})}{\sum_{m' \in M^{SP}} \exp(\frac{V_m^{SP}}{\theta_d^{SP}})} \tag{18}$$

The log-likelihood function of the RP and SP combined model is:

$$\ln(L^{RP+SP}) = \ln(L^{RP}) + \ln(L^{SP}) \tag{19}$$

where $\ln(L^{RP})$ and $\ln(L^{SP})$ are the log-likelihood functions for RP and SP data respectively, which are defined as follows:

$$\ln L^{RP} = \sum_{n=1}^{N^{RP}} \sum_{d=1}^{D^{RP}} \sum_{m=1}^{M^{RP}} \ln\left\{(P_{nm|d}^{RP})^{\delta_m^{RP}} (P_{nd}^{RP})^{\delta_d^{RP}}\right\} \tag{20}$$

$$\ln L^{SP} = \sum_{n=1}^{N^{SP}} \sum_{d=1}^{D^{SP}} \sum_{m=1}^{M^{SP}} \ln\left\{(P_{nm|d}^{SP})^{\delta_m^{SP}} (P_{nd}^{SP})^{\delta_d^{SP}}\right\} \tag{21}$$

where $\delta_m^{RP}(\delta_m^{SP})$ and $\delta_d^{RP}(\delta_d^{SP})$ are the dummy variables for an individual $n$ choosing mode $m$ to travel to destination $d$ in the RP (SP) choice context.

*3.2 The Trip Generation Model*

The Poisson regression models are employed for modeling trip generation. Individual characteristics and the accessibility index, which were calculated based on the RP/SP combined destination and mode choice model, are used for explanatory variables.

In the Poisson regression model, the numbers of trips are modeled as a Poisson random variable with a probability of occurrence being

$$P(Y_i) = \frac{e^{-\lambda} \lambda^{Y_i}}{Y_i!} \tag{22}$$

where $Y_i$ is individual $i$'s number of trips, and $\lambda$ is parameterized as follows:

$$\ln(\lambda) = \Sigma \varsigma_l G_l \tag{23}$$

where $\varsigma_l$ is the parameter of $G_l$ which is the $l$-th explanatory variable.

It is well-known from literature as well as from practice that with the use of the accessibility measure for the expected maximum utility of the individual from origin zone $i$ to destination zone $j$

$$Accessibility_i = \ln\left(\sum_{D_i} \exp(V_j^{SP})\right) \tag{24}$$

Where $D_i$ is the destination alternatives set for origin $i$, and $V_j^{SP}$ is the utility function for choosing destination $j$ ($j \in D_i$).

Thus, the proposed nested structure allows us to capture the influence of adding new travel modes and improving level of services on trip generation via the measurement of accessibility, i.e., the induced travel. Note that the model can only capture the short-term induced travel. To capture the long-term effects of adding new travel modes and improving level of services, changes in land use, such as residential location and facility location, need to be modeled.

## 4. DATA

A comprehensive questionnaire with both Revealed Preference (RP) and Stated Preference (SP) questions was designed for conducting a survey. The questionnaire includes two main parts: In the RP section, respondents were asked to fill in a trip diary of all inter-city trip generation in the past one year; whereas, the SP part was carefully tailored to collect travelers intended mode and destination choices, when HSR is available.

The survey, which was conducted in 2011, covered all of the Hanoi (Vietnam) area including the urban-core, urban ring, and sub-urban areas. Respondents were randomly selected in the designed places and interviewers conducted face-to-face surveys. People who have inter-city travel experience in the North-South corridor were selected to answer a full set of the questionnaire, whereas the others were only asked for their social demographic characteristics.

In the RP part, respondents were asked to fill in the trip information depending on trip purposes, including origin and destination, travel cost, travel time, access and egress time, and so on. Information about the travel party was also asked for non-business trips. In the SP survey, the interviewer first introduces and describes the HSR to the respondent. Then the respondent was given a set of questionnaires in which the hypothetical attributes of each travel mode, such as travel cost, travel time, and service frequency, were systematically varied.

The destinations along the proposed HSR route are divided in seven zones because of its geography and tourism characteristics. Explanatory variables include zone characteristics as well as the logsum variable of the maximum utility of the mode choice nest.

The available modes in the current situation are inter-city buses, conventional rails, conventional airlines, LCCs, and cars. In the SP design, the car is not included in the choice set for long distance (LD) trips. The choice set of SP surveys for business trip of middle and long trip distances includes conventional airline, LCC, and HSR since the assumption of travel cost is paid by the travelers' organizations. For the non-business trips of SP surveys, HSR was added to the choice set among all other current available modes.

The summary of survey data and summary of data characteristics are shown in Table 1 and Table A (in Appendix A), respectively.

Table 1. Summary of survey data

|  |  | Business | | Non-business |
|---|---|---|---|---|
|  |  | Mode choice | Destination choice | Mode choice & Destination choice |
| Number of trip | RP | 407 | 407 | 446 |
|  | SP | 2432(608) | | 1216(608) |
| Number of individual |  | Destination choice | | Destination choice |
|  |  | 246 | | 524 |

*Note:* The numbers of respondents are given in parentheses

## 5. MODEL ESTIMATION AND RESULTS

*5.1 Mode Choice and Destination Choice Models*

The estimation results of business trip and non-business trip models are shown in Table 2 and Table 3, respectively. We can confirm that the goodness-of-fits (i.e., Rho-squared) are sufficiently high, and the parameters of most explanatory variables and logsum variables are statistically significant at the level of 95%, and all the parameters have expected signs. The detailed discussions are given below.

### 5.1.1 The business trip model estimation results

In this study, the destination is defined as a group of neighboring provinces/cities (See Appendix B). Due to the lack of secondary data, only the total GRP at destination is used as the explanatory variable for destination choice. The estimated coefficient is statistically significant and represents the influence of GRP level on the destination choice of the business traveler.

The scale parameter for RP data model is less than one, which means that the SP data has more noise than that of the RP data.

Regarding the travel mode choice, the results show that the travel cost, in-vehicle travel time, access and egress time have negative influences, which mean that the cheaper the price and the shorter the travel time, the more likely users are to select that travel mode. All of those parameters are statistically significant. All the dummy variables representing the influence of RP on SP are significant, which are consistent with previous studies. This implies that the choices of mode in SP are highly influenced by the actual behavior. For the constant terms, the negative parameters for bus and conventional rail indicate that the businessperson has a preference to travel by car if all other variables are the same. However, the opposite situation applies for airline and HSR as those parameters have positive signs. It is intuitively understandable that the value of access and egress time is bigger than that of in-vehicle travel time.

The Occupation (1 if Government officer, and Office staff, 0 otherwise), Education (1 if Have university degree or above, 0 otherwise), and Car ownership (1 if Owning car, 0 otherwise) were used as explanatory variables for theta. For young government officers or office staff and business people who have university degrees or above and have higher income, the mode choice is strongly influenced by the destination choice of business trips. Car owners have negative influence as they are more independent with regards to their travel modes.

### 5.1.2 The non-business trip model estimation results

It can be seen from the estimation results that the parameter of attractiveness of destination, which is measured by the number of tourists visited the place, is statically significant. It can be inferred that tourists are likely to travel to destinations with more tourist arrivals, which can be explained by the effects of social interaction. Respondents may think that the destinations visited by higher numbers of tourists will be more attractive than others. The parameters of the dummy variables for summer vacation as seasonal influence vary due to the variance of tourist attraction in each zone. It is found that tourists are more likely to travel to Zones 3, 4, and 7 where there are famous beaches. Perhaps, tourists may be more interested in temperate places as the respondents' location is in Hanoi where it is very hot in the summer time. Zones 2 and 5, which are famous for traditional and historical tourism, have negative influences on travelers' destination choice during the summer break time. The experience or perception of respondents about tourist attractions in the current situation has considerable impact on the SP destination choice. In other words, travelers' knowledge about present tourism resources statistically

Table 2  Estimation results of mode choice and destination choice for business trip

| Explanatory variables | | Parameter (significant level) | | | | | |
|---|---|---|---|---|---|---|---|
| **Destination choice** | | | | | | | |
| Log of Total GRP (10^6Mil VND) | RP | 0.572 (**) | | | | | |
| **Theta explanatory variables** | | | | | | | |
| Occupation*Age | RP | -0.069 (**) | | | | | |
| Education level*Income | RP | 1.198 (**) | | | | | |
| Car ownership | RP | -0.284 (**) | | | | | |
| Constant | RP | 11.675 (**) | | | | | |
| **Mode choice** | | | | | | | |
| Travel cost (Mil VND) | All | -2.189 (**) | | | | | |
| In-vehicle travel time (mins) | All | -1.65e-3 (**) | | | | | |
| Access and egress time (mins) | All | -3.45e-3 (**) | | | | | |
| Influence of RP on SP mode choice | | | | | | | |
|   Airline | SP | 20.402 (**) | | | | | |
|   LCC | SP | 28.204 (**) | | | | | |
| Constant | | Bus | CR | Car | Airlines | LCC | HSR |
|   Zone 2 | RP | -2.527 (**) | -2.978 (**) | 0 | - | - | - |
|   Zone 3 | RP | -2.359 (**) | -2.665 (**) | 0 | - | - | - |
|   Zone 4 | RP | -0.986 (**) | -1.602 (**) | 0 | - | - | - |
|   Zone 5 | RP | -1.189 (**) | -1.935 (**) | 0 | -1.089 (**) | - | - |
|   Zone 6 | RP | -7.565 (**) | -1.755 (**) | 0 | 2.864 (**) | -2.537 (**) | - |
|   Zone 7 | RP | -0.673 (**) | -0.947 (*) | - | 4.402 (**) | 0 | - |
|   Zone 8 | RP | 1.729 (**) | 2.765 (**) | - | 4.412 (**) | 0 | - |
|   All zone | SP | - | - | - | 6.167 (**) | 0 | 2.944 (**) |
| **Scale parameter for RP data** | | 0.239 (**) | | | | | |
| LL0 | | -4043.026 | | | | | |
| LL1 | | -2826.928 | | | | | |
| rho | | 0.3007 | | | | | |
| rho.adj | | 0.2928 | | | | | |
| VOT (in-vehicle time) | All | 45,304.59 | | | | | |
| VOT (Access and egress time) | All | 94,725.56 | | | | | |
| Number of observation | | 928 | | | | | |

*(\*) significant at 90% level, (\*\*) significant at 95% level*

influences the location of future travel destinations. In addition, the results show that tourists who visited Zones 6 and 8 are likely to come again, which can be clarified by a variety of tourist attractions in those areas. Tourists can see temples, castles, historical sites, heritage sites, ancient tombs, and other attractions remaining from the past as well as some annual festivals, as well as beautiful beaches and resorts in Zone 6. On the other hand, apart from the scores of historical places, natural attraction sites, Zone 8 is famous for being the center for shopping, dining and relaxation.

    The destination choices of high-income people traveling with family are more affected by the level of services of travel modes, while working people show the opposite tendency.

    For the mode choice model, it is found that the travel cost, and in-vehicle travel time have negative influences on travel mode choice. The parameters that represent the influence of RP on SP mode choice have positive signs and are statistically significant, meaning that the travelers tend to repeat their choice behavior with the modes they have experienced. This empirical study provides evidence of the influence of stated preferences on revealed preferences. The negative parameters of income influence on bus and conventional rail users indicate that tourists with the same income have preferences to choose the car if other variables are the same. Whereas the travelers with higher income are likely to choose airline or HSR for

Table 3 Estimation results of mode choice and destination choice for non-business trip

| Explanatory variables | | Parameter (significant level) | | | | | |
|---|---|---|---|---|---|---|---|
| **Destination choice** | | | | | | | |
| Attractiveness of destination (Millions) | RP | 0.014 (**) | | | | | |
| Seasonal vacation (1 if Summer (May, June, July), 0 otherwise): | | | | | | | |
|   Zone 2 | RP | -0.390 (**) | | | | | |
|   Zone 3 | RP | 0.785 (**) | | | | | |
|   Zone 4 | RP | 0.911 (**) | | | | | |
|   Zone 5 | RP | -0.376 (**) | | | | | |
|   Zone 6 | RP | 0.181 | | | | | |
|   Zone 7 | RP | 0.414 (**) | | | | | |
| Tourism attraction (RP factor) | SP | 0.297 (*) | | | | | |
| Influence of RP on SP destination choice | | | | | | | |
|   Zone 6 | SP | 8.333 (**) | | | | | |
|   Zone 8 | SP | 1.662 (**) | | | | | |
| **Theta explanatory variables** | | | | | | | |
| Marital Status (1 if Married, 0 otherwise) | All | 0.246 | | | | | |
| Age | All | -0.011 | | | | | |
| Income*Travel party (1 if Travel with family, 0 otherwise) | All | 0.132 (**) | | | | | |
| Working status (1 if Working, 0 otherwise) | RP | -0.404 (**) | | | | | |
| Constant | RP | 1.845 (**) | | | | | |
| Constant | SP | 0.633 (**) | | | | | |
| **Mode choice** | | | | | | | |
| Travel cost (Million VND) | All | -1.073 (**) | | | | | |
| In vehicle travel time (mins) | All | -8.48e-4 (**) | | | | | |
| Influence of RP on SP mode choice | | | | | | | |
|   Bus | SP | 5.961 (**) | | | | | |
|   Conventional Rail | SP | 6.741 (**) | | | | | |
|   Airline | SP | 4.572 (**) | | | | | |
|   LCC | SP | 7.514 (**) | | | | | |
| Income (Mil VND) | | Bus | CR | Car | Airlines | LCC | HSR |
|   Zone 2,3,4 | RP | -0.075 (**) | -0.042 (**) | 0 | - | - | - |
|   Zone 5 | RP | -0.075 (**) | -0.042 (**) | 0 | 0.179 (**) | - | - |
|   Zone 6 | All | -0.075 (**) | -0.042 (**) | 0 | 0.183 (**) | 0.076 (**) | - |
|   Zone 7 | RP | -0.075 (**) | -0.042 (**) | - | 0.104 (**) | 0 | - |
|   Zone 8 | All | -0.075 (**) | -0.042 (**) | 0 | 0.090 (**) | -0.014 | 0.142 (**) |
| Constant | | Bus | CR | Car | Airlines | LCC | HSR |
|   Zone 2 | RP | -0.700 (**) | -2.964 (**) | 0 | - | - | - |
|   Zone 3 | RP | -0.518 (**) | -1.267 (**) | 0 | - | - | - |
|   Zone 4 | RP | -0.591 (**) | -0.958 (**) | 0 | - | - | - |
|   Zone 5 | RP | -0.821 (**) | -1.015 (**) | 0 | -3.563 (**) | - | - |
|   Zone 6 | RP | 0.608 (**) | 0.813 (**) | 0 | -1.056 (**) | -1.769 (**) | - |
|   Zone 7 | RP | -0.011 | 0.465 (**) | - | 0.797 (**) | 0 | - |
|   Zone 8 | RP | 1.129 (**) | 1.229 (**) | - | 0.978 (**) | 0 | - |
|   Zone 6 | SP | 1.424 (**) | 1.600 (**) | 0 | 1.602 (**) | 0.412 | 1.467 (**) |
|   Zone 8 | SP | -0.428 | -0.121 | 0 | 2.185 (**) | -0.576 (**) | -0.814 (**) |
| **Scale parameter for RP data** | | 0.672 (**) | | | | | |
| LL0 | | -5053.843 | | | | | |
| LL1 | | -3697.605 | | | | | |
| rho | | 0.2683 | | | | | |
| rho.adj | | 0.2562 | | | | | |
| VOT (in-vehicle time) | All | 47,423.11 | | | | | |
| Number of observation | | 880 | | | | | |

*(*) significant at 90% level, (**) significant at 95% level*

their nonbusiness trips. All constant terms of short distances and middle distances (MDs) have negative signs, which indicate that tourists prefer to travel by car if all other variables are the same. On the other hand, airline has a higher priority to be chosen for long trip distances.

The scale parameter for the RP and SP data model is less than one, which means that the RP data has less noise than that of SP data.

It is found that the value of in-vehicle travel time for business trips is smaller than that of for non-business trips. This result contradicts other current literature and needs to be investigated further since there are no references to such types of studies in Vietnam.

*5.2 Trip Generation Models*

Only resident-based respondents in RP data were selected for estimating the trip generation by using the Poisson regression model. The results are shown in Table 4 and Table 5. The goodness-of-fit indicators (Pseudo-R square) are sufficiently high. All parameters have expected signs and statistically significant at the level of 90% or 95%. The detailed discussions are given below.

Table 4. Estimation results of trip generation for business trip

| Explanatory variables | Parameter | Std. Error | t value | Pr(>|z|) (significant level) |
|---|---|---|---|---|
| (Intercept) | -4.200 | 1.990 | -2.111 | 0.035 (**) |
| Occupation | | | | |
| Businessperson (base) | 0 | - | - | - |
| Government officer/ office staff | -0.959 | 0.128 | -7.498 | 6.49e-14 (**) |
| Industrial laborers | -0.948 | 0.332 | -2.854 | 0.004 (**) |
| Others | -0.527 | 0.210 | -2.506 | 0.012 (**) |
| Education level | | | | |
| Senior high school (base) | 0 | - | - | - |
| College/Vocational training | -0.546 | 0.287 | -1.902 | 0.057 (*) |
| Bachelor | 0.778 | 0.222 | 3.502 | 0.001 (**) |
| Master/doctor | 1.083 | 0.276 | 3.928 | 8.58e-05 (**) |
| Others | 1.143 | 0.272 | 4.196 | 2.72e-05 (**) |
| Age | 0.011 | 0.006 | 2.048 | 0.041 (**) |
| Accessibility | 0.603 | 0.306 | 1.970 | 0.049 (**) |
| Pseudo-R square | 0.172 | | | |
| Chi-square (significant level) | 245.64 (<0.001) | | | |
| Number of sample | 246 | | | |

*(\*) significant at 90% level, (\*\*) significant at 95% level*

Table 5. Estimation results of trip generation for non-business trip

| Explanatory variables | Parameter | Std. Error | t value | Pr(>|z|) (significant level) |
|---|---|---|---|---|
| (Intercept) | -2.078 | 0.243 | -8.564 | < 2e-16 (**) |
| Education (Having university or above degree: 1; otherwise: 0) | 0.224 | 0.105 | 2.122 | 0.033 (**) |
| Married status (Married:1; otherwise: 0) | 0.187 | 0.111 | 1.693 | 0.090 (*) |
| Gender (Male: 1; female: 0) | -0.174 | 0.096 | -1.814 | 0.069 (*) |
| Accessibility | 1.321 | 0.175 | 7.554 | 4.22e-14 (**) |
| Pseudo-R square | 0.188 | | | |
| Chi-square (significant level) | 157.52 (<0.001) | | | |
| Number of sample | 524 | | | |

*(\*) significant at 90% level, (\*\*) significant at 95% level*

### 5.2.1 The business trip generation model estimation results

The estimation results indicate that business people are likely to produce more trips than respondents with other occupations. Respondents who hold college or vocational training or university or above degrees seem to generate more business trips than people who graduated from senior high school, which can be explained their knowledge of the benefits of business trips. People who had finished college/vocational training produce less business trips than persons who have senior high school diplomas since the explanatory parameter is negative and significant at 90%. This may need to be studied further.

The respondents with higher age are statically significant on the production of business trips. This seems that the older person will have more experience and can be valued more in business meeting.

As the accessibility is calculated from equation (24), any change in the level of service can influence the trip production. For the business trip, accessibility has a positive sign and is significant which means that business people are likely to increase his travel frequency to the zones with higher accessibility. For the utility function used to estimate the accessibility of business trip production, the total GRP of the zone was used, hence, the more developed the areas are the more business trips it will attract.

The significant constant means that some other explanatory variables, which have influence on the business trip production, are not included in this model. In order to clearly understand the business production behavior pattern, it is essential to get a better understanding of the unobserved factor.

### 5.2.2 The non-business trip generation model estimation results

The results of the non-business trip generation model show that the person has a higher education level, is more likely to create more nonbusiness trip. In addition, the married couple has positive and significant influence on the tourism trip generation since they have a partner to travel with, the trips will be more interesting because of communications between partners during the trips. The negative parameter of the male traveler might be because of family commitments since the man in Vietnam is considered a primary breadwinner as head of the family and hence would have lower probability to participate in tourism.

The parameter of accessibility is positive and statistically significant as expected. It implies that non-business travelers seem more likely to travel to the zones, which are easier to access. Moreover, the total tourists visiting a corresponding zone and summer time were used as explanatory variables for the accessibility. More people visited may suggest the easier of accessibility to the place. In addition, many local governments may well prepare for infrastructure before the peak tourism time as well as travel agencies are likely to offer some discounts in summer time, hence, the place may become more attraction since tourists are given better options to travel.

*5.3 Simulation Analysis*

To ensure the vision for the introduction of HSR and before the enormous investment for HSR infrastructure and the implementation of the operation format, strategic scenarios are necessary to simulate, define and analyze the various possibilities. In general, the competition in modal share of inter-city travel is mainly between HSR, conventional airline, and LCC with the service levels, such as travel time, and travel cost. Additionally, the HSR system has its own sets of advantages, like being fast, safe and comfortable. Also due to economic growth and higher incomes, the value of time may become more important and hence the ridership of the future HSR may be expected to increase. In Table 6, the scenarios are the simplified versions of reality in the planning year, for the brief outlook of the introduction of HSR. In a Report on Construction Investment of HSR Line Hanoi – Ho Chi Minh City – Final Report of the Vietnam and Japan Consultant Joint Venture (VJC), profiles of HSR's service levels are set up to observe the modal share among HSR, conventional airline, and LCC with the assumption that other attributes of other alternatives do not change. The sensitivity of travel demand was examined by single and multiple factors of service level. For the first five scenarios, the competition of HSR and airlines can be observed. Then, HSR technology can be examined by changing of its travel time. The low cost carrier may be common in the future, therefore, scenario 8 to 10 was setup to inspect the competition of low cost carrier with HSR and airlines. The influence of location of HSR station and accessibility to airport were tested in scenario 11 to scenario 14. Finally, the development level among zones was studied in scenario 15.

Simultaneous simulations were done for MD and LD mode choice and destination choice. Due to the scores of scenarios, the results will be presented separately for MD, LD, and destination choice with regards to the trip purposes.

For business trip, the simulation results show from Figure 4 a, b, c. When travel cost of HSR is lower than that of airline, HSR is the dominant mode of MD while the airline is the dominant mode of LD. By comparing with the current mode share of business trip in Figure 2, HSR share mostly seizes from car and inter-city bus, and conventional airlines for MD and LD respectively. Through all scenarios, the share in destination choice of business trip between MD and LD is quite stable at around 50%, however when the development among Zones was set up equally, the significance change in choice of MD and LD can be observed.

For non-business trips, the similar results with business trips can be seen. The HSR is the dominant mode of MDs while that of LDs is airlines. In addition, the cheaper price of HSR and LCC are, the longer travel distance is. However, the share of future HSR for MDs is the shifted from car, bus and conventional rail, while for LDs, bus and conventional rail travelers are likely to change to HSR. The details of simulation results for non-business purpose can be seen from Figure 4 d, e, and f.

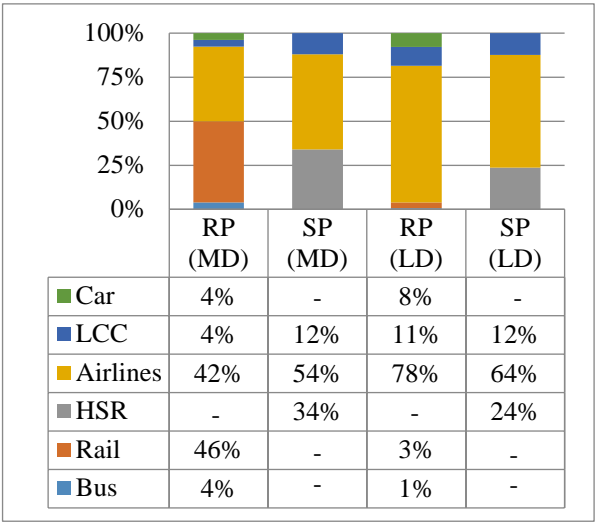

Figure 2 RP and SP mode share by MDs and LDs of business trips

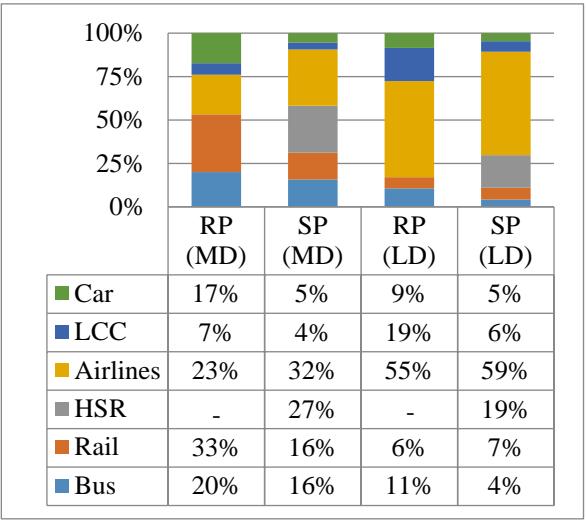

Figure 3 RP and SP mode share by MDs and LDs of non-business trips

### 5.4 Induced Travel

It is important to examine the influence on the trip generation when introducing new mode in

Table 6 Details of scenarios

| | Case | Explanations | Business | Non-business | Simulation meaning |
|---|---|---|---|---|---|
| Simulation for single factor | Scenario 1 (S1) | HSR travel cost equals 70% with that of airlines | o | o | Competition of HSR and airlines |
| | Scenario 2 (S2) | HSR travel cost equals 80% with that of airlines | o | o | |
| | Scenario 3 (S3) | HSR travel cost equals 90% with that of airlines | o | o | |
| | Scenario 4 (S4) | HSR travel cost equals with that of airlines | o | o | |
| | Scenario 5 (S5) | HSR travel cost equals 110% with that of airlines | o | o | |
| | Scenario 6 (S6) | HSR travel time equals 80% as planed | o | o | HSR technology |
| | Scenario 7 (S7) | HSR travel time equals 120% as planed | o | o | |
| | Scenario 8 (S8) | LCC travel cost equals 50% with that of airlines, HSR | o | o | Competition of LCC, HSR and airlines |
| | Scenario 9 (S9) | LCC travel cost equals 70% with that of airlines, HSR | o | o | |
| Simulation for multiple factors | Scenario 10 (S10) | HSR travel cost equals 70% with that of airlines and HSR travel time equals 80% as planed  LCC travel cost equals 50% with that of airlines | o | o | |
| | Scenario 11 (S11) | HSR travel cost equals 70% with that of airlines and HSR travel time equals 80% as planed  HSR station is located in city center  (Access and egress time decrease 30% for HSR) | o | × | Location of HSR station |
| Simulation for multiple factors | Scenario 12 (S12) | HSR travel cost equals 70% with that of airlines and HSR travel time equals 80% as planed  HSR station is located outside city center  (Access and egress time increase 30% for HSR) | o | × | Location of HSR station |
| | Scenario 13 (S13) | HSR travel cost equals 70% with that of airlines and HSR travel time equals 80% as planed  HSR station is located outside city center  Access and egress time decrease 25% for Airline | o | × | |
| | Scenario 14 (S14) | HSR travel cost equals 70% with that of airlines and HSR travel time equals 80% as planed  HSR station is located outside city center  Access and egress time decrease 50% for Airline | o | × | Location of HSR station and accessibility to airport |
| | Scenario 15 (S15) | Same with scenario 14 and GDP of zone 6 equals to that of zone 8. | o | × | Development level |

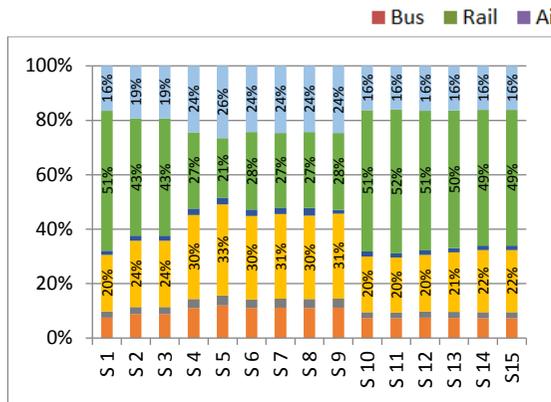

a) Mode choice of MD for business trips

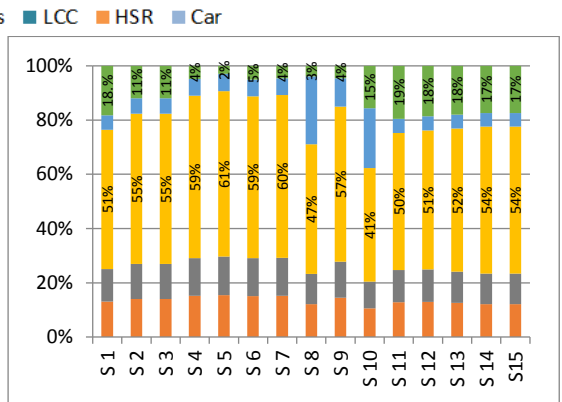

b) Mode choice of LD for business trips

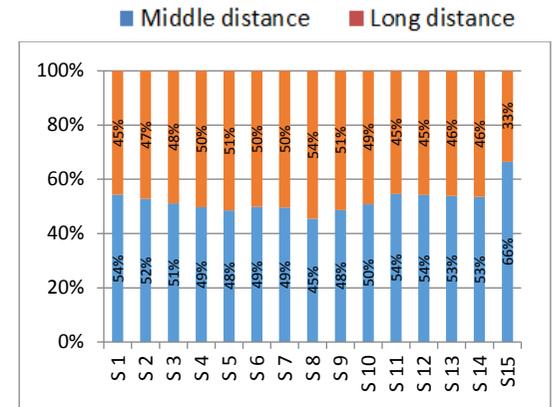

c) Destination choice for business trips

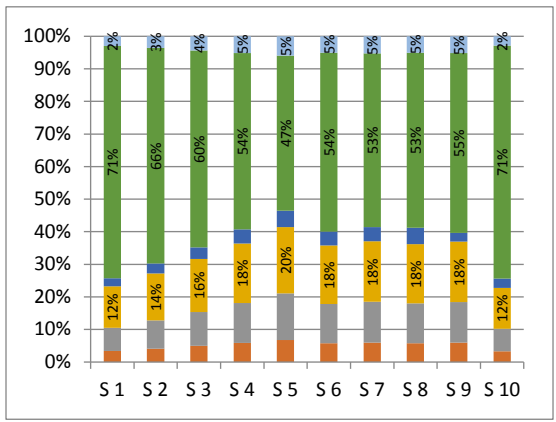

d) Mode choice of MD for non-business trips

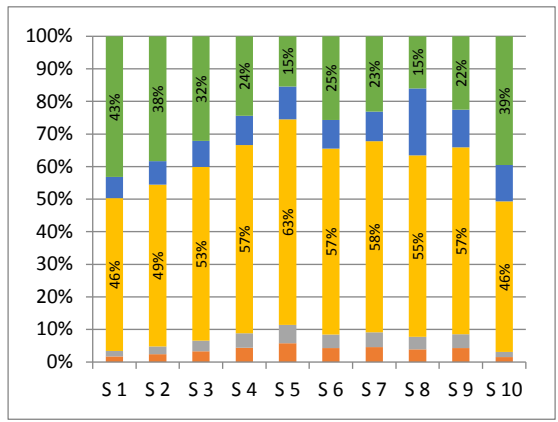

e) Mode choice of LD for non-business trips

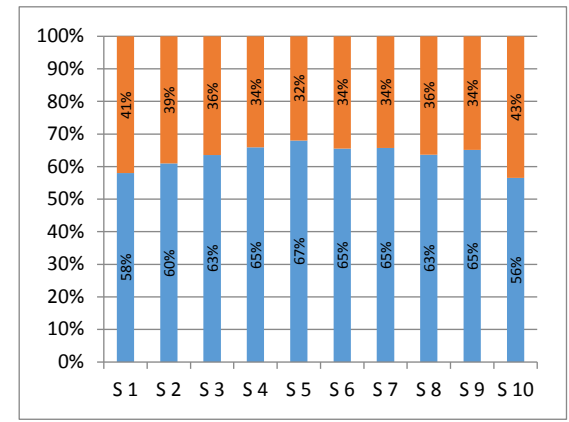

f) Destination choice for non-business trips

Figure 4. Simulation results

Table 7. Induced travel for business and non-business trips

|  | S1 | S2 | S3 | S4 | S5 | S6 | S7 | S8 |
|---|---|---|---|---|---|---|---|---|
| **Business** | 1.468 (100.89) | 1.465 (100.69) | 1.461 (100.41) | 1.455 (100.00) | 1.453 (99.86) | 1.458 (100.21) | 1.452 (99.79) | 1.459 (100.27) |
| **Non-business** | 0.901 (106.75) | 0.874 (103.55) | 0.858 (101.66) | 0.844 (100.00) | 0.832 (98.58) | 0.866 (102.61) | 0.842 (99.76) | 0.882 (104.50) |
|  | S9 | S10 | S11 | S12 | S13 | S14 | S15 | - |
| **Business** | 1.457 (100.14) | 1.476 (101.44) | 1.478 (101.58) | 1.471 (101.10) | 1.469 (100.96) | 1.472 (101.17) | 1.486 (102.13) | - |
| **Non-business** | 0.869 (102.96) | 0.926 (109.72) | - | - | - | - | - | - |

*The numbers in parenthesis are the percentages by compare to the base case (scenario 4 - S4)*

transportation market. In this paper, the sensitivity of trip generation can be captured by computing the changes in accessibility in each scenario, which is set up in previous part. The means of trip frequency of all scenarios were compared with that of scenario 4 as it is set as a base case.

Even the authors' model estimation results indicate that business trip is sensitive with the travel time, there is no significant influence of accessibility on trip generation through 15 scenarios. On the other hand, the larger sensitivity can be observed for non-business trip for both travel time and travel cost. This is understandable, since they have to manage travel by their own budget implying that the trip frequency is more sensitive with the changes of the travel cost. The change of multiple factors seems to have significant effects on the trip frequency of non-business purpose. On the other hand, it should be noted that the analysis made in this paper focuses only on changes in individuals' preferences through various travel cost and time, and does not consider relocations of facilities such as firms and factories. The latter impact would be significant in long term and it needs to be explored in future work.

## 6. CONCLUSION

Since a dramatically increasing travel demand along the North-South corridor in Vietnam is expected in the future, it might be necessary to have a better transportation system such as HSR to accommodate the future demand. The purpose of this research is to develop an integrated inter-city demand-forecasting model incorporating trip generation, destination choice, and mode choice for business and non-business trip purposes. A comprehensive survey, which includes both RP and SP questions were conducted in Hanoi in 2011. Apart from travel cost and travel time, which influences the mode choice for any purpose, income has a statistically significant impact on the mode choice of non-business trips. Trip frequency, accessibility and individual characteristics were employed to estimate the trip generation model, whereas zonal and transportation characteristics were the input variables for the destination choice models. As a result, the influence of the introduction of HSR on changes in destination choice as well as mode choice behavior can be observed. The study successfully established the integrated inter-city travel demand models, which can be captured the induced travel for business and non-business trip purposes. Model estimation results show that individual characteristics, destination specific attributes, and travel specific attributes significantly influence the three choices (trip frequency, destination choice and mode choice) of travelers for both business and non-business trips. Finally, simulation results reveal that HSR may be the dominant mode for MDs while for LD trips it is the conventional airlines. It was also found that with lower fares from HSR and LCCs, travelers tend to travel longer distances when compared with conventional airlines.

The induced travel reveals that the trip frequency is likely to increase when the levels of services have improvement, such as the shorter in access time and in-vehicle travel time, and the cheaper of travel cost, the more attractive of destination. The results also indicate that non-business trips are more sensitive with the levels of services than business trips. The important future task is to explore the impacts of HSR on facility relocations, which will also affect the travel demand.

## ACKNOWLEDGMENT

This research was mainly supported by budget of the Japanese Government for Japanese Grant Aid for Human Resource Development Scholarship (JDS) program, and partly by the Global Environmental Leaders (GELs) Education Program of Hiroshima University, Japan. In addition, we would also like to thanks Dr. Sudarmanto Budi Nugroho for his comments and suggestions during manifold meetings of this study.

## APPENDIX A

Table A. Summary of data characteristics

| Individual characteristics | Percentage (%) | Individual characteristics | Percentage (%) |
|---|---|---|---|
| **Age** | | **Final academic degree** | |
| 18,19 | 6.03 | Senior high school | 14.51 |
| 20-29 | 37.36 | College/Vocational training | 18.61 |
| 30-39 | 28.37 | Bachelor | 56.61 |
| 40-49 | 12.07 | Master / Doctor degree | 5.01 |
| 50-59 | 10.01 | Others | 5.26 |
| ≥60 | 6.16 | **Income** (million VNĐ) | |
| **Gender** | | ≤ 1,600 | 11.04 |
| Male | 53.15 | 1,601 - 3,000 | 19.38 |
| Female | 46.85 | 3,001 - 5,000 | 21.57 |
| **Occupation** | | 5,001 - 10,000 | 27.98 |
| Government officer/ Office staff | 37.48 | 10,001 - 15,000 | 10.53 |
| Industrial laborer | 7.83 | 15,001 - 20,000 | 6.55 |
| Merchant | 7.83 | > 20,000 | 2.95 |
| House-wife/ Jobless/ Retied | 11.04 | **Marital status** | |
| Student, pupil | 18.87 | Single | 35.94 |
| Others | 16.94 | Married | 61.49 |
| | | Others | 2.57 |

# APPENDIX B

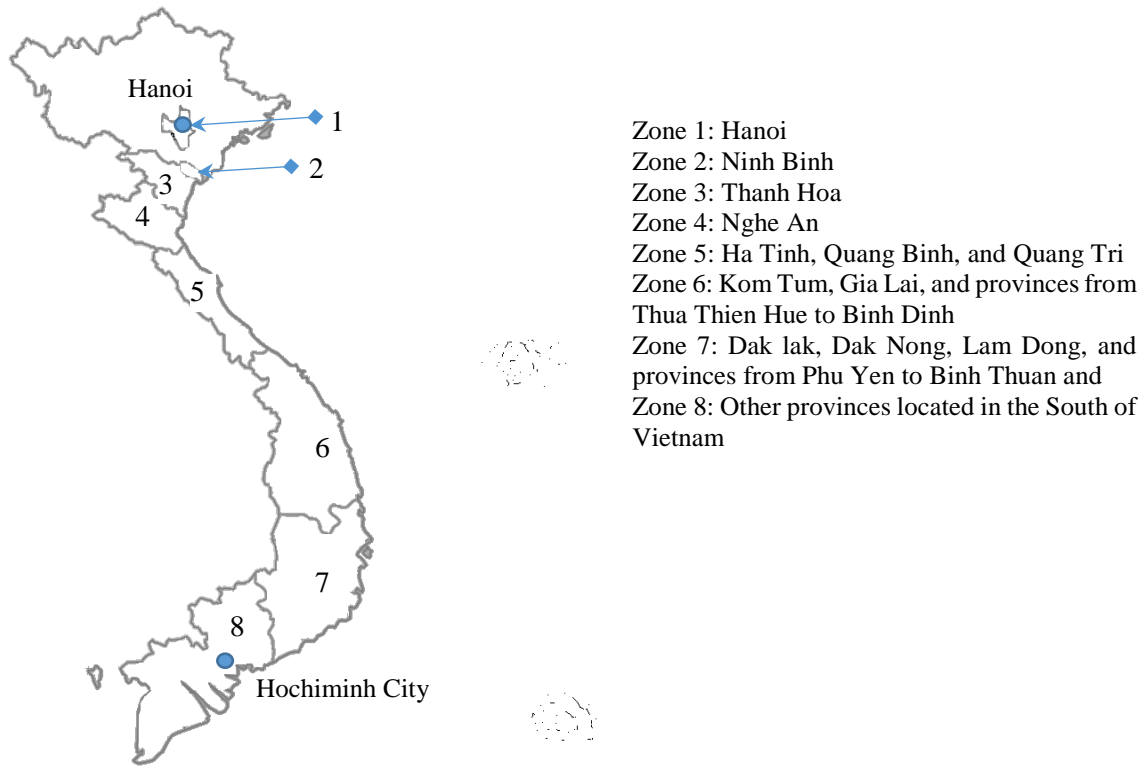

Zone 1: Hanoi
Zone 2: Ninh Binh
Zone 3: Thanh Hoa
Zone 4: Nghe An
Zone 5: Ha Tinh, Quang Binh, and Quang Tri
Zone 6: Kom Tum, Gia Lai, and provinces from Thua Thien Hue to Binh Dinh
Zone 7: Dak lak, Dak Nong, Lam Dong, and provinces from Phu Yen to Binh Thuan and
Zone 8: Other provinces located in the South of Vietnam

Figure B. Map of destination alternatives